\documentclass[preprint,prd,aps,showpacs,showkeys,onecolumn,english]{revtex4}
\usepackage{amsmath,amssymb} 
\usepackage[dvips]{graphicx} 
\usepackage{graphics}
\usepackage{graphicx}
\usepackage{epsfig}
\usepackage{slashed}
\usepackage[english]{babel}

\def\bentarrow{\:\raisebox{1.3ex}{\rlap{$\vert$}}\!\longrightarrow}

\newcommand\nn{\nonumber}
\newcommand\ba{\begin{eqnarray}}
\newcommand\ea{\end{eqnarray}}

\newcommand{\GeV}{~\mbox{GeV}}

\begin{document}
\title{A study of the process $e^+ +e^- \to e^+ +e^- +p + \bar {p}$ by the two-photon mechanism $\gamma \gamma \to p \bar {p}$ at high energies}
\author{Azad~I.~Ahmadov$^{a,b}$ \footnote{E-mail: ahmadov@theor.jinr.ru}}
\affiliation{$^{a}$ Bogoliubov Laboratory of Theoretical Physics,
JINR, Dubna, 141980 Russia}
\affiliation{$^{b}$ Institute of Physics, Azerbaijan
National Academy of Sciences, H.Javid ave. 131, AZ-1143 Baku, Azerbaijan}

\date{\today}

\begin{abstract}
In this paper, we consider the exclusive production of proton-antiproton pairs in the interaction between two quasireal
photons in $e^+e^-$ collision.
The differential and total cross section of the process $\gamma \gamma \to p\bar {p}$ at a beam energy of photons
from 2.1 GeV to 4.5 GeV in the center-of-mass and for different values of $|cos\theta^{\ast}|$ is calculated.
At energy $<\sqrt {s_{e^+e^-}}>$ = 197 \,\,GeV the total cross section process of the $e^+ +e^- \to e^+ +e^- +p +\bar {p}$
is calculated by the two-photon mechanism. The results are in satisfactory agreement with the experimental data.

\vspace*{0.5cm}

\noindent
\end{abstract}

\maketitle

\section{Introduction}
\label{Introduction}
As before, one of the main problems in the microworld physics is the confirmation of the standard model and the search
for possible exits beyond its limits \cite{salam1,salam2,salam3}.
Therefore, $e^+ +e^- \to e^+ +e^-, \,\,e^+ +e^- \to \mu^+ +\mu^-, \,\,
e^+ +e^- \to \tau^+ +\tau^-, \,\,e^+ +e^- \to q +\bar{q}$, è $e^+ +e^- \to e^+ +e^- +p +\bar {p}$
reactions are the main processes studied at electron-positron colliders.
In studying these processes by the L3, OPAL, ALEPH, and DELPHI Collaborations in the LEP experiments no
deviations from the predictions of the Standard Model (SM) were observed  \cite{a2,a3,a4,a5,a6,a7,a8,a9,a10,a11,a12,a13}.

At the present time, an experiment conducted at the LHC $pp$-beams with energies of 14 TeV in the center-of-mass system
gives a unique opportunity for comparing different predictions of the SM in this energy range.
At high energies new experimental facts, may appear which are not consistent with the predictions of the SM.

High-energies of electron-positron colliders are an appropriate object for studying the
two-photon process $e^+ e^- \to e^+ e^- \gamma^{\star} \gamma^{\ast} \to e^+ e^- X$, where $\gamma^{\ast}$ denotes a virtual photon.
It should be noted that this process is an extensive source of hadrons.
In this reaction, the outgoing electron and positron beam carry almost the full energy, and usually due to their small transverse momenta.
The final state of $X$ has, therefore, a small mass compared to the $e^+e^-$ center-of-mass energy $\sqrt{s}$ and the transverse momentum
is almost zero.

Small virtuality of photons allows one to extract the cross section $\sigma(\gamma \gamma \to X)$ for real photon collisions,
which are calculated by the QED \cite{budnev}.

For the critical test of perturbative QCD experiment of proton-antiproton pair production at $e^+e^-$ colliders was investigated
at LEP energies \cite{a13,L3}.
The study of two-photon processes $\gamma\gamma \to X$ represents an important section of the modern high-energy physics \cite{budnev,morgan}.
Traditionally, they are studied in experiments with the $e^+e^-$ colliding beams at the interaction of virtual photons, that are
emitted by the initial particles (that is $e^+ e^- \to e^+ e^- \gamma^{\ast} \gamma^{\ast} \to e^+ e^- X$).
Is this case, a system of particles $X$ with invariant mass of $W$ is formed, and the cross-section of this process in the equivalent photon
approximation has the form
\ba
d\sigma^{e^+ e^- \to e^+ e^- X}(s) = dn_1 dn_2 d\sigma^{\gamma\gamma \to X}(W^2),
\ea
where $dn_i$ are the spectra of equivalent photon emission:
\ba
dn_i = \frac{\alpha}{\pi}\frac{d\omega_i}{\omega_i}\frac{d(-q_i^2)}{(-q_i^2)}\biggl[1-\frac{\omega_i}{E}+
\frac{\omega_i^2}{2E^2}+\frac{m^2\omega_i^2}{q_i^2E^2}\biggr].
\ea
Here $E$ is the energy in the center-of-mass system of an electron (positron) beam, $m$ is the electron mass,
$\omega_i$ and $q_i$ are the energy and four-momentum of the virtual photon ($i$=1,2), $W=\sqrt{4\omega_1\omega_2}$.

To check the predictions of QCD the exclusive production of the proton-antiproton pairs ($p\bar {p}$) in the collision
of two quasireal photons is studied.
This process has been studied in a LEP experiment.
In the LEP experiment, the photons are emitted by electron beams and $p\bar {p}$ pairs are produced in the process
$e^+ e^- \to e^+ e^- \gamma^{\ast} \gamma^{\ast} \to e^+ e^- p\bar {p}$ \cite{telnov}.

Quantum chromodynamics has been used for predicting cross-sections for exclusive hadron pairs of high transverse momentum
in the collision of two photons.

General theory of hard exclusive processes in QCD is studied in detail in Refs.\cite{ch1,ch12,lep,lepb,ch2},
and at high energies and fixed center-of-mass angle, an analytic expression for the differential cross section of the
process of $\gamma \gamma \to p\bar {p}$ is obtained.

One of these processes for the study of the perturbative QCD is hadron production in the final state by photon-photon interactions.
Therefore, many authors \cite{far1,an,kr,mil,hy,kr1} performed in the developed framework \cite{lep,lepb} and
presented calculations of the total cross section for $\gamma \gamma \to p\bar {p}$.

Using the wave function of the proton on the basis of the QCD sum rules \cite{ch3}, the first estimate of the cross section for the process
$\gamma \gamma \to p\bar {p}$  was obtained in the three-quark system ($n_c$=8) \cite{far1,far2}.
For example, in the diquark model \cite{an,kr,ber} the proton is considered as a quark-diquark system.

Other approaches, for example, the bags model \cite{di} was developed for large transfer momentum, and calculations were applied
at intermediate energies ($W_{\gamma\gamma} >$2.55 GeV).
In recent years calculation in the perturbative QCD was successfully applied to describe many of the inclusive scattering processes
with large transfer momentum \cite{ven}.
In addition, in exclusive hadron scattering one can find cross-sections in good agreement with the power energy
dependence predicted by QCD \cite{br,brf,mmt},
for transmitting momentum $|t| > 5 \,\,GeV^2$, for example, in the experimental data on nucleon-nucleon scattering,
Compton scattering, photoproduction and form factors of the pion and proton.
The cross sections for production of proton-antiproton pairs in $\gamma\gamma$ collisions were also calculated in
perturbative QCD \cite{far1,br1,dam}.

The hadron production in the interaction of two photons is the study of the process

\ba
\gamma^{\ast} + \gamma^{\ast} \to {\rm hadrons}
\ea
at fixed virtual photons $q_i^2 = -Q_i^2 <0 $, and as for large squares of the center-of-mass energy $W^2 = (q_1 + q_2)^2$ with
$q_i$ of photon momenta.
In the present case, if one considers a high limit for $W$, the cross-section behaves as expected as follows:

\ba
\sigma_{\gamma^{\ast}\gamma^{\ast}} \sim \frac{1}{W^2}.
\ea
A simple way to study this process is through the reaction,
\ba
\begin{array}{rcl}
e^+ + e^- & \longrightarrow & e^+ + e^- + \underbrace{\gamma^* + \gamma^*} \\
 &  & \phantom{e^+ + e^- + \gamma^*\:}\bentarrow {\rm hadrons} ;
\end{array}
\label{hadrons}
\ea
namely, $e^+e^-$ collisions, are considered selecting those events in which incoming leptons produce two photons, which eventually
initiate hard scattering, that is hadron production.

It is clear, the multitude of Feynman diagrams contributes to the process, which is really observed
\ba
e^+ + e^- \to e^+ + e^- + {\rm hadrons}.
\ea
Investigation of production of proton-antiproton pairs in $\gamma
\gamma$ collisions at high energy is one of the most interesting
problems for the  phenomenology of the $\gamma \gamma$ collisions.

In this study, we apply the two photon mechanism to compute  the exclusive production
of proton-antiproton pairs in the interaction between two quasireal photons in electron-positron collision.

We are interested in the calculation and analysis of the dependence of differential
and total cross sections on the center-of-mass energy for different scattering angles
and also angular distribution of the differential cross-section of the
process $e^+ e^- \to e^+ e^- \gamma^{\ast} \gamma^{\ast} \to e^+ e^- p\bar {p}$  and finally
results are compared with experimental ones.

In this paper, we present the studying of the differential and total cross-sections of the process $\gamma \gamma \to p\bar {p}$
in the energy region 2.1 GeV$<W_{\gamma\gamma}<$4.5 GeV and different values of $cos\theta^{\ast}$,
where $\theta^{\ast}$ is the scattering angle of the proton or antiproton relative to the direction of the
incoming photons in the $\gamma\gamma$ center-of-mass system of the reaction $e^+ +e^- \to e^+ e^- p \bar {p}$.
We also study the angular dependence and the energy behavior of the cross-section of the process $\gamma \gamma \to p\bar {p}$,
and calculate the total cross-section of the process $e^+ +e^- \to e^+ e^- p \bar {p}$ by the two-photon mechanism.

Experimental results of the proton-antiproton pair production by two photons collisions in processes $e^+ +e^- \to e^+ e^- p \bar {p}$
and $\gamma \gamma \to p\bar {p}$ were presented in Refs. \cite{a12,a13}.

The paper is planned as follows. The formulae for the calculation differential and total cross-sections of the  process
$\gamma \gamma \to p\bar {p}$ are provided in Section \ref{ht}.
Some formulae and analysis of the cross-section of the process $e^+ +e^- \to e^+ e^- p \bar {p}$ the dependence
on the center-of-mass energy and the angular distributions within two-photon mechanism  approach is presented in Section
\ref{ir}. Finally, some concluding remarks are stated  in Section \ref{conc}.

\section{The process $\bf {\gamma\gamma \to p \bar{p}}$} \label{ht}

\begin{figure}[!htb]
       \centering
       \includegraphics[width=0.8\linewidth]{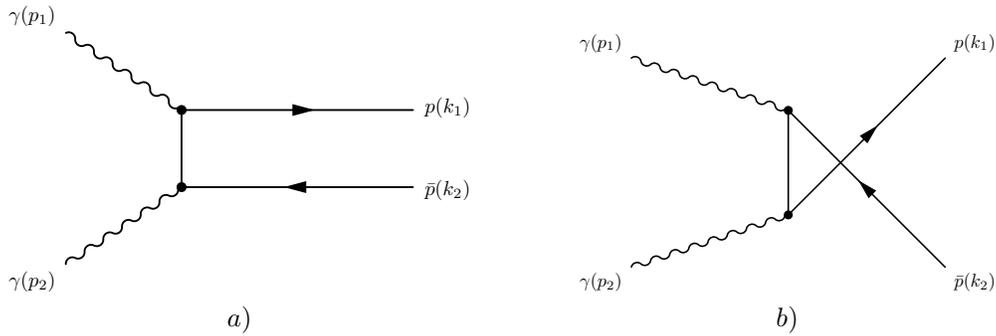}
       \caption{Feynman diagrams for the production of $p \bar {p}$ in $\gamma \gamma$ collisions}
       \label{prod1}
\end{figure}

Here, we are on process to develop this direction, for this reason, at the first stage we will consider the
proton-antiproton pair production in $\gamma \gamma$ collisions

\ba
\gamma + \gamma \to p + \bar{p}.
\label{gg}
\ea
The Feynman diagrams process \eqref{gg} is shown in Fig.~\ref{prod1}

The amplitude of the process \eqref{gg}, which is depicted by two Feynman diagrams (see Fig.~\ref{prod1}), can be written as follows:

\ba
{\mathcal{M}}_a = -e^2\bar{u}(k_1)\Gamma_{\mu}^t (k_1, k_1-p_1) \frac{\slashed {p}_2 - \slashed {k}_2 +M_p}{(k_2-p_2)^2-M_p^2}\Gamma_{\nu}^t (k_1-p_1, -k_2) v(k_2)
\varepsilon_{\mu}(p_1)\varepsilon_{\nu}(p_2),  \nn \\
{\mathcal{M}}_b = -e^2\bar{u}(k_1)\Gamma_{\nu}^u (k_1, k_1-p_2) \frac{\slashed {k}_1 - \slashed {p}_2 +M_p}{(p_2-k_1)^2-M_p^2}\Gamma_{\mu}^u (k_1-p_2, -k_2) v(k_2)
\varepsilon_{\mu}(p_1)\varepsilon_{\nu}(p_2),
\label{M}
\ea
where $M_p$ is the proton mass, $k_1$ and $k_2$ are the proton and antiproton momenta,  \\
$\varepsilon_{\mu}(p_1),\,\,\varepsilon_{\nu}(p_2)$ are the polarization vectors of the  photons
with initial state momentum $p_1$ and $p_2$, respectively.

Here, $\Gamma_{\mu (\nu)}^{t(u)}$ is the photon-proton vertex functions (form factors), which are determined in the following form:
\ba
\Gamma_{\mu (\nu)}^t = \gamma_{\mu (\nu)}F_1(t)+\frac{i}{2M_p}\sigma^{\mu(\nu)\rho}(p_1 - k_1)^{\rho}F_2(t), \nn \\
\Gamma_{\mu (\nu)}^u = \gamma_{\mu (\nu)}F_1(u)+\frac{i}{2M_p}\sigma^{\mu(\nu)\rho}(k_2 - p_1)^{\rho}F_2(u), \nn \\
\sigma^{\mu\rho} = \frac{i}{2}(\gamma^{\mu}\gamma^{\rho} - \gamma^{\rho}\gamma^{\mu}),
\label{Gam}
\ea
Here, $F_1(t),\,\,F_2(t),\,\,F_1(u)$ and $F_2(u)$ are the Dirac and Pauli proton form factors for the $t$ and $u$ channels, respectively.  \\
Following Refs.\cite{FF1,FF2}, the form factors $F_1(t)\,\,F_2(t),\,\,F_1(u)$ and $F_2(u)$ can be parametrized as:
\ba
F_1(t) = \frac{\Lambda^4}{\Lambda^4 +(t-M_p^2)^2},\,\,\,\,\,F_2(t) = k_p F_1(t), \nn \\
F_1(u) = \frac{\Lambda^4}{\Lambda^4 +(u-M_p^2)^2},\,\,\,\,\,F_2(u) = k_p F_1(u),
\label{formfactor}
\ea
which, in the real photon limit, coincide with the static values $F_1(M_p^2)=1,\,\,F_2(M_p^2)=k_p$.
$k_p=1.798$ is the anomalous magnetic moment of the proton with respect to the coupling with photon, and
$\Lambda =0.911$ is an empirical cutoff.

If it is required to produce summation over the photon polarization, it is necessary to replace
$\varepsilon_{\mu}(p_1,\lambda)\varepsilon_{\nu}^{\ast}(p_1,\lambda)$ by the expression

\ba
\sum_{\lambda}\varepsilon_{\mu}(p_1,\lambda)\varepsilon_{\nu}^{\ast}(p_1,\lambda) = -g_{\mu\nu}.
\ea
The cross section of the process \eqref{gg} is parameterized in terms of the following Mandelstam variables:
\ba
s = (p_1 + p_2)^2 = (k_1 + k_2)^2;  \nn \\
t = (p_1 - k_1)^2 = (p_2 - k_2)^2;  \nn \\
u = (p_1 - k_2)^2 = (p_2 - k_1)^2.
\label{Man}
\ea
They are connected, as one can easily, by the relation
\ba
s + t + u = 2M_p^2.
\label{Man1}
\ea
By considering the fact that  the real photon doesn't have  mass, then we
obtain:

\ba
p_1^2 = p_2^2 =0,
\label{Man2}
\ea

Then the square of the  full amplitude has the form:

\ba
\overline{|{\mathcal M}|^2} = \frac{1}{4}\sum({\mathcal{M}}_a {\mathcal{M}}_a^{\ast} + {\mathcal{M}}_b {\mathcal{M}}_b^{\ast} +
2Re({\mathcal{M}}_a {\mathcal{M}}_b^{\ast})).
\label{M2}
\ea
The differential cross section of the process \eqref{gg} after spin
averaging  can be written in the following form:
\ba
\frac{d\sigma}{dt} = \frac{1}{16 \pi s^2} \frac{k_{c.m.}}{p_{c.m.}}\overline{|{\mathcal M}|^2},
\label{ds}
\ea
where $k_{c.m.}=\frac{1}{2}\sqrt{s-4M_p^2},\,\,\,  p_{c.m.}=\frac{1}{2}\sqrt{s}$.  \\
For obtaining  the total cross section of the process \eqref{gg} we
use the following formula:
\ba
\sigma (s) = \int\limits_{t^-}^{t^+} dt \frac{d\sigma}{dt}.
\label{cs}
\ea
For a given $\sqrt s$ energy of center-of-mass system the relation
between $t$ and $u$ defined by the border equation in this form

\ba
t u = M_p^4,
\label{tu}
\ea

Then for  given value of the  center-of-mass energy and together solving of these two equations \eqref{Man1} and
\eqref{tu} we can find upper and lower bounds of the integral \eqref{cs} in this form:
\ba
t^{\pm} = \frac{-s+2M_p^2 \pm \sqrt{s(s-4M_p^2)}}{2}.
\label{tpm}
\ea
Using the trace techniques, the squared amplitudes \eqref{M2}
explicitly take the following form, we calculate the square of the
amplitudes \eqref{M2}, and  instead of $u$  make the substitution  $u = 2M_p^2 - s - t$, from Eq. \eqref{Man1} and then we get
\newpage
%
\ba
\overline{|{\mathcal M}|^2} &=& \frac{1}{\frac{1}{4}(M_p^4 - 2M_p^2t + t^2)}\biggl[F_1(t)F_2^3(t)\biggl(144 M_p t^3 - 288 M_p^3 t^2 + 144 M_p^5 t \biggr) + \nn \\
&&+F_1^2(t) F_2^2(t) \biggl(- 8 M_p^2 s t + 36 M_p^2 t^2 + 60 M_p^4 t - 52M_p^6 + 8 s t^2 - 44 t^3 \biggr) + \nn \\
&&+ F_1^3(t) F_2(t) \biggl(48 M_pt^2 - 96M_p^3 t + 48 M_p^5 \biggr) + F_1^4(t) \biggl(2M_p^2 s - 6 M_p^4 - 2s t - 2t^2 \biggr) + \nn \\
&&+ F_2^4(t) \biggl(8 M_p^2 s t^2 + 48 M_p^2 t^3 + 24 M_p^4 t^2 - 32 M_p^6 t - 8 s t^3 - 40t^4\biggr)\biggr] +  \nn \\
&&+\frac{1}{\frac{1}{4}(M_p^4 - 2M_p^2 s -M_p^2 t +2st +s^2 + t^2)}\biggl[F_1^2(u)F_2^2(u) \biggl(240 M_p^2 st + 108 M_p^2s^2 + \nn \\
&&+132 M_p^2t^2 -164 M_p^4s - 180 M_p^4t + 76M_p^6 - 76st^2 - 68s^2t - 20s^3 - 28t^3 \biggr) +  \nn \\
&&+F_1^4(u) \biggl(6M_p^2s + 8M_p^2t - 14M_p^4 - 2st - 2t^2 \biggr)+ F_2^4(u) \biggl(776 M_p^2st^2 + 736 M_p^2s^2t + \nn \\
&&+232 M_p^2s^3 + 272M_p^2 t^3 - 1232 M_p^4st - 584 M_p^4s^2 - 648 M_p^4t^2 + 608 M_p^6s + 640 M_p^6t - \nn \\
&&-224 M_p^8 - 152 st^3 - 216 s^2t^2 - 136 s^3 t - 32 s^4 - 40 t^4 \biggr)+ \nn \\
&&+\frac{1}{\frac{1}{4}(-M_p^4 + 2M_p^2 s +M_p^2 t -st  - t^2)}
\biggl[F_1(t)F_2(t)F_1^2(u) \biggl(40M_p st + 32M_p t^2 - 24 M_p^3 s -  \nn \\
&&-64 M_p^3 t + 32 M_p^5 \biggr) + F_1(t)F_2(t)F_2^2(u) \biggl(112 M_p st^2 + 80 M_p s^2 t + 32 M_p t^3 -
-304 M_p^3 st - \nn \\
&&-64 M_p^3 s^2 - 144 M_p^3 t^2 + 192 M_p^5s + 192 M_p^5t - 80 M_p^7 \biggr)+ F_1^2(t)F_1^2(u) \biggl(4 M_p^2s - 16 M_p^4 \biggr) + \nn \\
&&+F_1^2(t)F_2^2(u) \biggl(36 M_p^2 st + 12 M_p^2 s^2 + 24 M_p^2 t^2 - 28 M_p^4 s - 36 M_p^4 t + 16 M_p^6 - 8 st^2 - \nn \\
&&-4s^2t - 4t^3 \biggr) + F_2^2(t) F_1^2(u) \biggl(4 M_p^2 st + 12 M_p^4t - 8 M_p^6 - 4st^2 - 4t^3 \biggr) + \nn \\
&&+ F_2^2(t) F_2^2(u) \biggl(- 32 M_p^2st^2 -32M_p^2s^2t + 112M_p^4 st + 48M_p^4t^2 - \nn \\
&&-80M_p^6s - 96M_p^6t + 48M_p^8 \biggr)\biggr].
\label{M22}
\ea

\begin{figure}[!htb]
       \centering
       \includegraphics[width=0.7\linewidth]{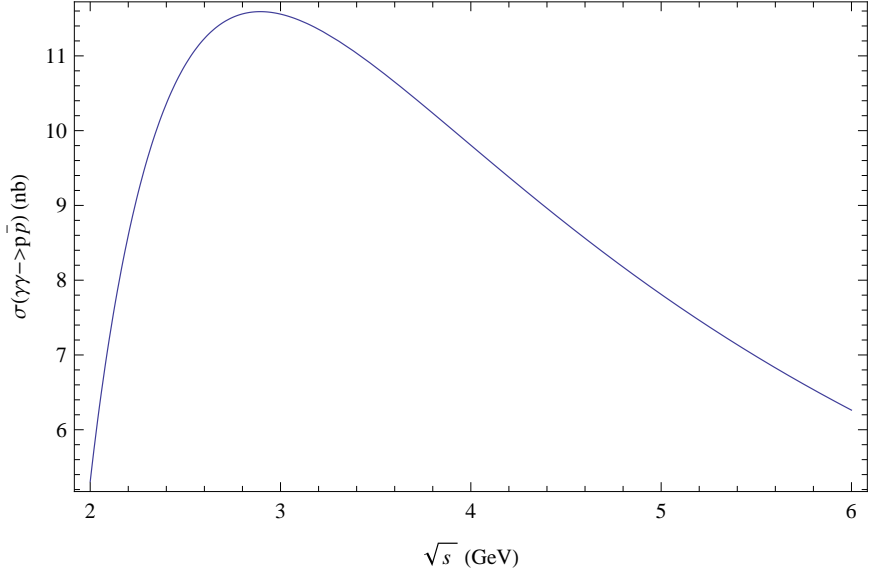}
       \caption{The total cross section of the process $\gamma \gamma \to p\bar {p}$ as a function of the CM energy.}
       \label{sigma1}
\end{figure}

\begin{figure}[!htb]
       \centering
       \includegraphics[width=0.7\linewidth]{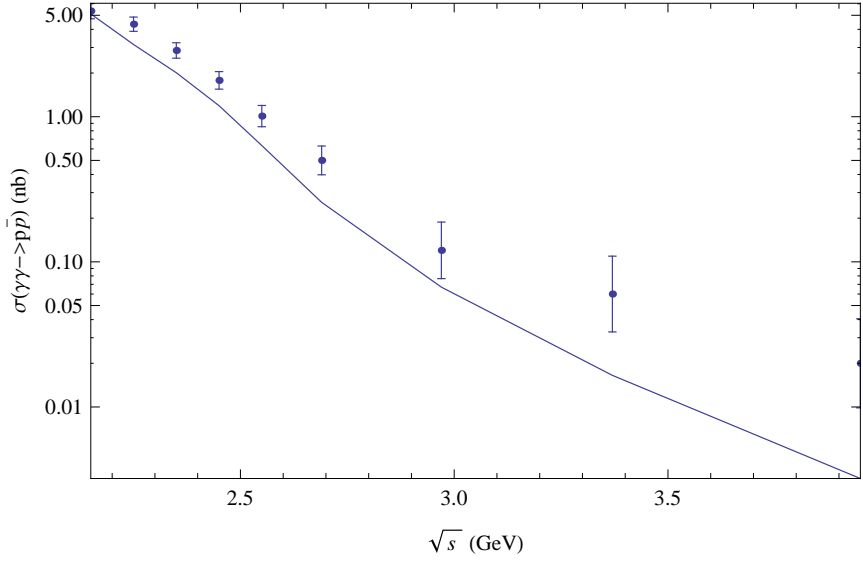}
       \caption{The total cross section of the process $\gamma \gamma \to p\bar {p}$ as a function of the CM energy for the large angle region, $|cos\theta^{\ast}| < 0.6$
       is compared with the experimental data \cite{a12}
             with the total center-of-mass energy $\sqrt{s} = 2.1 \div 4.5~\GeV$.}
       \label{sigma2}
\end{figure}

\begin{figure}[!htb]
       \centering
       \includegraphics[width=0.7\linewidth]{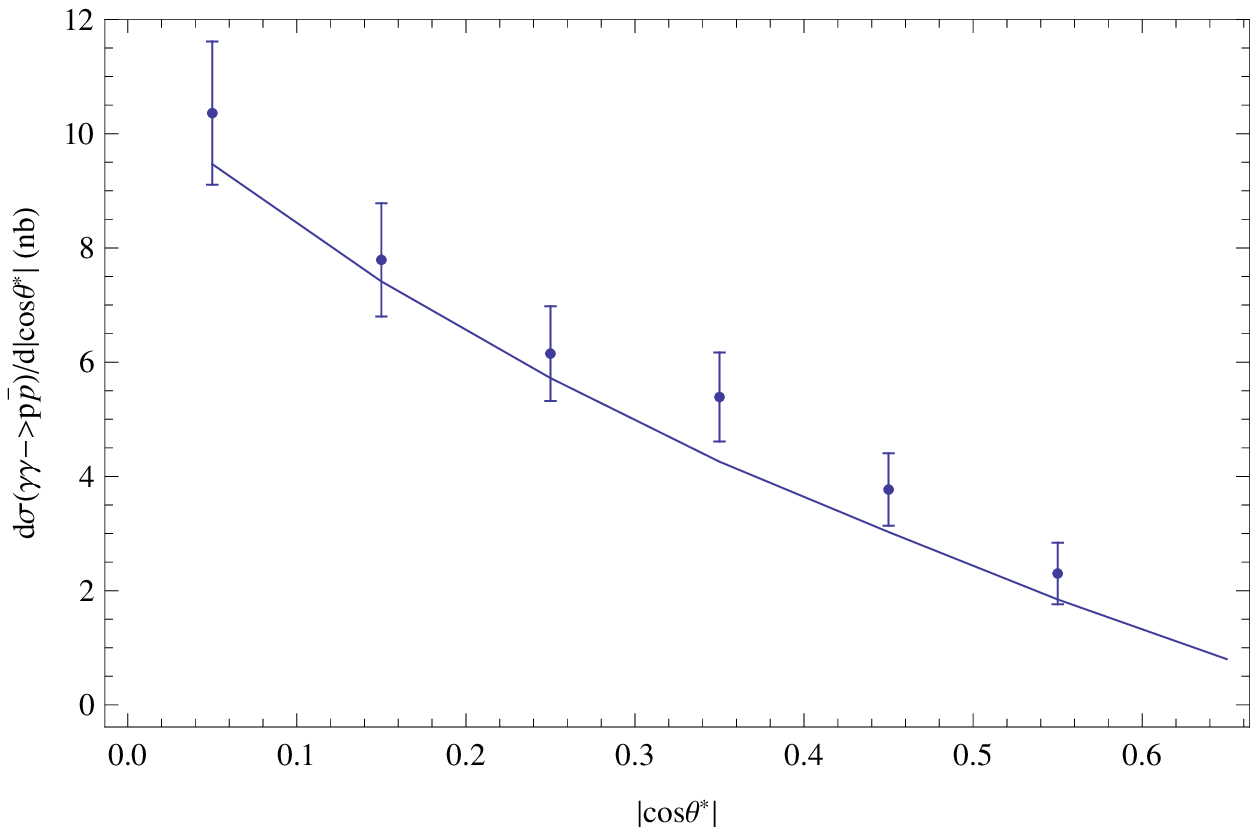}
       \caption{The differential cross section of the process $\gamma \gamma \to p\bar {p}$ as a function of the $|cos\theta^{\ast}|$ for the energy region
       2.1 GeV$<W_{\gamma\gamma}<$2.5 GeV
       is compared with the experimental data \cite{a12}.}
       \label{dsigma31}
\end{figure}

\begin{figure}[!htb]
       \centering
       \includegraphics[width=0.7\linewidth]{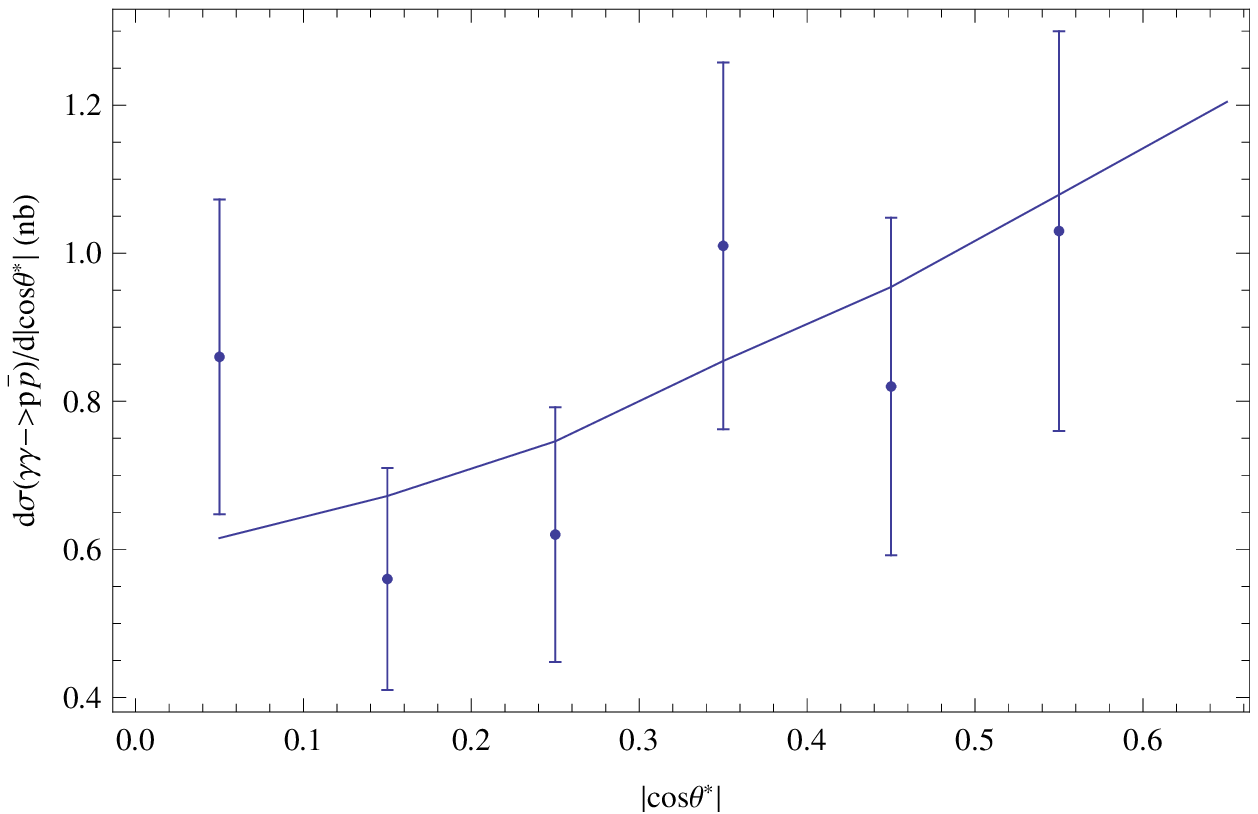}
       \caption{The differential cross section of the process $\gamma \gamma \to p\bar {p}$ as a function of the $|cos\theta^{\ast}|$ for the energy region
       2.5 GeV$<W_{\gamma\gamma}<$3.0 GeV
       is compared with the experimental data \cite{a12}.}
       \label{dsigma32}
\end{figure}

\begin{figure}[!htb]
       \centering
       \includegraphics[width=0.7\linewidth]{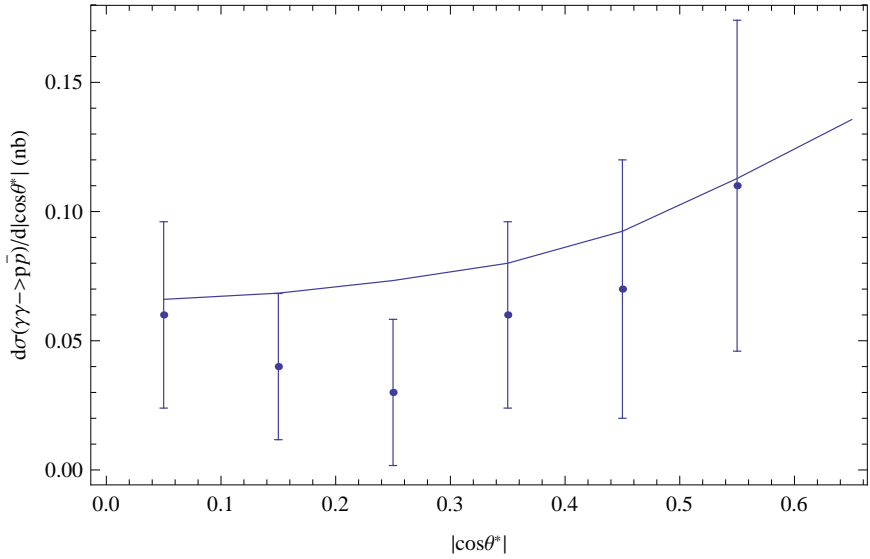}
       \caption{The differential cross section of the process $\gamma \gamma \to p\bar {p}$ as a function of the $|cos\theta^{\ast}|$ for the energy region
       3.0 GeV$<W_{\gamma\gamma}<$4.5 GeV and
       is compared with the experimental data \cite{a12}.}
       \label{dsigma33}
\end{figure}

In the Eq.\eqref{M22} we use the following expressions:
\ba k_1^2 = k_2^2 = M_p^2, \,\,\,(p_1 p_2) = s/2; \,\,\,(k_1 k_2) =
s/2 - M_p^2; \,\,
(p_1 k_1) = (M_p^2 -t)/2;  \nn \\
(p_2 k_2) = (M_p^2 -t)/2; \,\,\,(p_1 k_2) = (M_p^2 -u)/2; \,\,\,(p_2
k_1) = (M_p^2 -u)/2.
\label{Man3}
\ea
We calculated the total cross section  of the  process \eqref{gg}
proton-antiproton production in $\gamma\gamma$ interactions
\eqref{gg} at center-of-mass energy from $\sqrt {s}$ = 2 GeV to 6
GeV.
Also we investigated the energy distribution of the total cross section of the  $\gamma \gamma \to p\bar {p}$ process for different value, of $|cos\theta^{\ast}|$. \\
In Fig.~\ref{sigma1} is shown  the dependence of the total cross
section of the process  $\sigma(\gamma \gamma \to p\bar {p})$ as a
function of the center-of-mass energy $\sqrt {s}$ .
In Fig.~\ref{sigma2} the energy dependence of the total cross
section of the process $\gamma \gamma \to p\bar {p}$ as a function
of the energy at $|cos\theta^{\ast}| < 0.6$ is presented.
The differential cross section for the angular distribution is expressed in the following form:
\ba
\frac{d\sigma}{dcos\theta^{\ast}} = \frac{(4\pi\alpha)^2}{32\pi s}\sqrt{1-\frac{4M_p^2}{s}}\cdot \overline{|{\mathcal M}|^2}.
\label{ds1}
\ea
The parameter $t$ is defined in the following form:
\ba
t = (p_1 - k_1)^2 = M_p^2 - \frac{s}{2} + \frac{\sqrt {s}}{2}\sqrt{s - 4M_p^2} cos\theta^{\ast}.
\label{t}
\ea
Now we will study the angular distribution of the differential cross section for the process \eqref{gg}
in different regions of the two-photon center-of-mass energy.
After integration over the energy $W_{\gamma\gamma}$ in the region of energy
2.1 GeV $< W_{\gamma\gamma} <$ 2.5 GeV,\,\,\,2.5 GeV$< W_{\gamma\gamma} <$ 3.0 GeV,\,\,\, 3.0 GeV $< W_{\gamma\gamma} <$ 4.5 GeV,
in the formulae \eqref{M22}, \eqref{ds1} and \eqref{t}, we get the formula for the differential cross section
as a function of $|cos\theta^{\ast}|$.
In Figs.~(\ref{dsigma31} - \ref{dsigma33}), we present separate dependences of the differential cross section on
$|cos\theta^{\ast}|$ at the two-photon energy 2.1 GeV $< W_{\gamma\gamma} <$ 2.5 GeV,\,\,\,2.5 GeV$< W_{\gamma\gamma} <$ 3.0 GeV,\,\,\, 3.0 GeV $< W_{\gamma\gamma} <$ 4.5 GeV, respectively.
Now we will study the  dependence of the total cross section on the
energy of the process \eqref{gg} in a large angle region
$|cos\theta^{\ast}|$. In formula \eqref{ds1} we carry out numerical
integration over $|cos\theta^{\ast}|$ using formulae \eqref{M22} and
\eqref{t} in the region of $|cos\theta^{\ast}| <$0.3, è 0.3 $<
|cos\theta^{\ast}| <$ 0.6. We obtain the total cross section of the
dependence on the two photons center-of-mass energy $\sqrt {s}$.

\begin{figure}[!htb]
       \centering
       \includegraphics[width=0.7\linewidth]{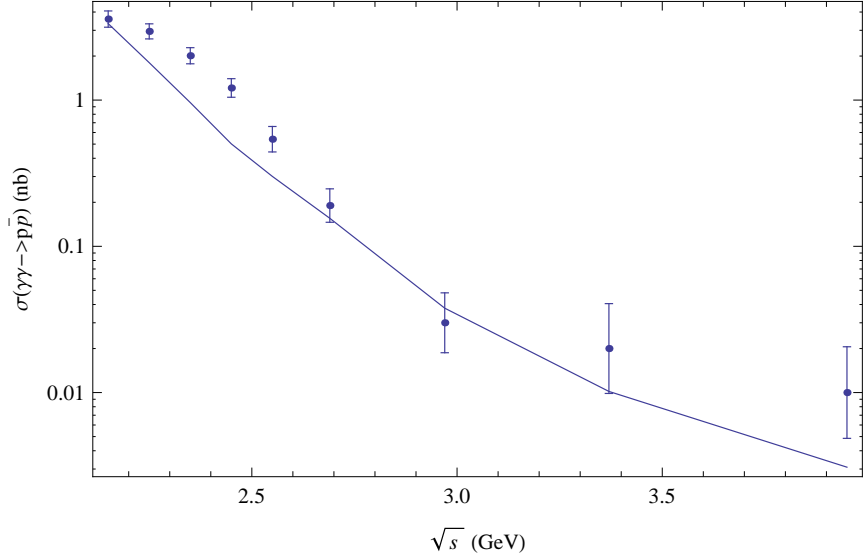}
       \caption{The  total cross section of the process $\gamma\gamma \to p \bar {p}$ as a function of $\sqrt {s}$ for the large angle region
       $|cos\theta^{\ast}| <$ 0.3,
       is compared with the experimental data \cite{a12}.}
       \label{dsigma34}
\end{figure}

\begin{figure}[!htb]
       \centering
       \includegraphics[width=0.7\linewidth]{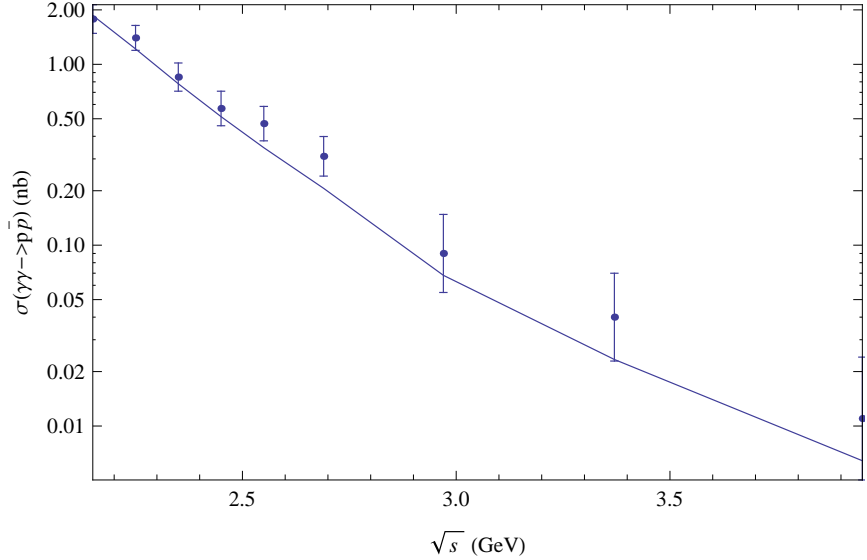}
       \caption{The total cross section of the process $\gamma\gamma \to p \bar {p}$ as a function of the $\sqrt {s}$ for the small angle region
       0.3 $< |cos\theta^{\ast}| <$ 0.6,
       is compared with the experimental data \cite{a12}.}
       \label{dsigma35}
\end{figure}

In Figs.~\ref{dsigma34} and ~\ref{dsigma35}  by using \eqref{ds1}, \eqref{M22},
and \eqref{t}  we present several types of
energy dependence  in region 2.1 GeV $< \sqrt {s} <$4.5 GeV of the
total cross section for $p \bar {p}$ pair production in$\gamma\gamma$ collision

In Figs.~\ref{dsigma34} and ~\ref{dsigma35} we present separate
energy dependence in region 2.1 GeV $< \sqrt {s} <$4.5 GeV of the
total cross section for $p \bar {p}$ pair production in
$\gamma\gamma$ collision on calculated by \eqref{ds1}, \eqref{M22},
and \eqref{t}.

Note that the results we have obtained for the process $\gamma\gamma \to p \bar {p}$ are in satisfactory agreement with the
known experimental data of the L3 Collaboration at LEP \cite{a12}.

For massive outgoing particles is the squared transverse momentum
\ba
p_{T}^2 = \frac{t u -M_p^4}{s},
\label{trans}
\ea
when masses are introduced in the final state.  \\
The Mandelstam invariants of the processes satisfy
\ba
t = (p_1 - k_1)^2 = (p_2 - k_2)^2 = M_p^2-\frac{X_{Tp}}{2}s e^{-y_p} = M_{\bar p}^2-\frac{X_{T{\bar p}}}{2}s e^{y_{\bar p}}, \nn \\
u = (p_1 - k_2)^2 = (p_2 - k_1)^2 = M_p^2-\frac{X_{Tp}}{2}s e^{y_p} = M_{\bar p}^2-\frac{X_{T{\bar p}}}{2}s e^{-y_{\bar p}}.
\label{tubys}
\ea
The transverse momentum $p_T$ distributions can be written in the following form
\ba
\frac{d\sigma}{dp_T^2} = \int dy_p \int dy_{\bar {p}} \cdot \frac{d\sigma}{dp_T^2 dy_p dy_{\bar {p}}},
\label{dpt}
\ea
the integration limits are
\ba
y_{\bar {p}min} = max\{\ln(\frac{x_{T\bar {p}}}{2-x_{Tp}e^{-yp}}); -Y_{\bar {p}}\}, \nn \\
y_{\bar {p}max} = min\{\ln(\frac{2-x_{Tp}e^{yp}}{x_{T\bar {p}}}); Y_{\bar {p}}\},
\ea
\ba
y_{pmax}=-y_{pmin} = min\{Y_p; \cosh^{-1}\biggl(\frac{1}{x_{Tp}}(1+\frac{M_p^2 -M_{\bar {p}}^2}{s})\biggr); \ln(\frac{2-x_{T{\bar {p}}}e^{-Y{\bar {p}}}}{x_{Tp}})\}.\,\,\,\,
\ea
The $x_{Tp}$ range would be
\ba
\frac{2M_p}{\sqrt {s}} \leq x_{Tp} \leq 1+ \frac{M_p^2 - M_{\bar {p}}^2}{s}.
\ea
Starting from this basic distribution, always assuming $M_p = M_{\bar {p}}$, and imposing the cuts
\ba
|y_p| = Y_p, \,\,\,\,|y_{\bar {p}}| = Y_{\bar {p}} \quad {\rm with} \quad Y_p = Y_{\bar {p}},
\ea
in the numerical applications we take $Y_{p,\bar {p}}=2$.

\vspace*{1.5cm}

%
\section{The process $\bf {e^+ + e^- \to e^+ + e^- + p + \bar{p}}$}\label{ir}

The two-photon mechanism of lepton and hadron production in
experiments with colliding electron-positron beams at high energies
was studied by several groups \cite{e1,e2,e3,e31,e4,e41,e5}. For the
production of the proton-antiproton pairs  within two-photon
mechanism we have the process of the type

\begin{figure}[!htb]
       \centering
       \includegraphics[width=0.5\linewidth]{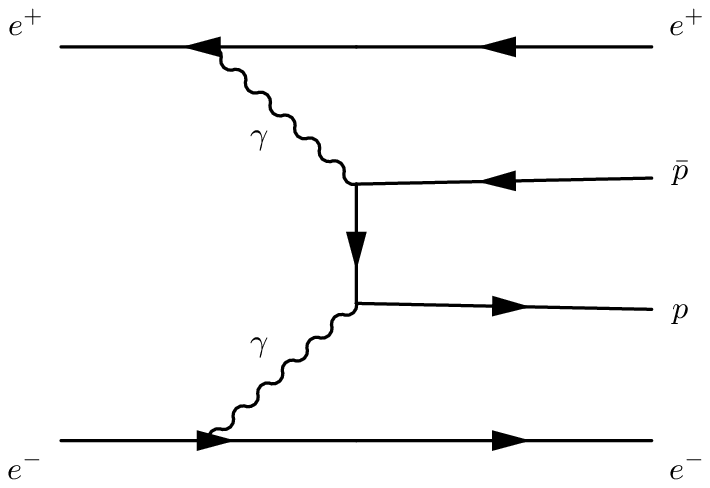}
       \caption{Feynman diagram of the process $e^+ + e^- \to e^+ + e^- + p + \bar{p}$.}
       \label{prod12}
\end{figure}

The two-photon mechanism of lepton and hadron production by colliding electron-positron beams at high energy
was studied experimentally  by several groups \cite{e1,e2,e3,e4,e5}. For the
production of the proton-antiproton pairs there exists the process of this type within two-photon
mechanism
\ba
e^+ + e^- \to e^+ + \gamma^{\ast} + e^- + \gamma^{\ast} \to e^+ + e^- + p + \bar{p}.
\label{ep}
\ea
The Feynman diagrams of the process \eqref{ep} are shown in Fig.~\ref{prod12}.
Also, in a particular case the production of heavy objects by means of two virtual photons in electron-positron collisions
was investigated by \cite{e1,e2,e6,e61,e62,e63,ak}.
The total cross section of the $e^+ + e^- \to e^+ + e^- + p + \bar{p}$ process can be written as follows \cite{budnev,e5}:
\ba
\sigma(s)^{e^+e^- \to e^+e^- p \bar{p}} = \biggl(\frac{\alpha}{\pi}\biggr)^2
\int\limits_{4M_p^2}^{4E^2}\frac{ds_1}{s_1}\sigma^{\gamma\gamma \to p \bar{p}}(s_1)\biggl[\biggl(\ln\frac{sM_p^2}{s_1 m_e^2}\biggr)^2 f\biggl(\frac{s}{s_1}\biggr)-
\frac{1}{3}\biggl(\ln\frac{s}{s_1}\biggr)^3\biggr],\,\,\,\,\,\,\,
\label{csee}
\ea
\ba
f(x) = \biggl(1+\frac{1}{2x}\biggr)^2\ln x - \frac{1}{2}\biggl(1-\frac{1}{x}\biggr)\biggl(3+\frac{1}{x}\biggr),
\label{f}
\ea
where $\sigma^{\gamma\gamma \to p \bar{p}}(s_1)$ is the cross section of the $\gamma\gamma \to p \bar{p}$ process.

Now we will study the dependence of the differential cross section
for the process $e^+ + e^- \to e^+ + e^- + p + \bar{p}$ on the
energy in a large angle region $|cos\theta^{\ast}|$. \\
The calculation of the differential cross section of the $e^+ + e^- \to e^+ + e^- + p + \bar{p}$ process by formulas \eqref{csee} is carried out in the energy region $\sqrt {s_{ee}}$ = 183--189 GeV for the OPAL Collaboration at LEP.
Also, in formula \eqref{csee} for $\frac{d\sigma^{\gamma\gamma \to p \bar{p}}(s_1)}{d|cos\theta^{\ast}|}$
in a numerical integration over $|cos\theta^{\ast}|$ in the region $|cos\theta^{\ast}| <$ 0.6 is carried out using formulae
\eqref{M22}, \eqref{ds1}, and \eqref{t}, and the two-photon center-of-mass energy $\sqrt {s_1}$ the  defined in 2.15 GeV $< \sqrt {s_1} <$3.95 GeV.

\begin{figure}[!htb]
       \centering
       \includegraphics[width=0.7\linewidth]{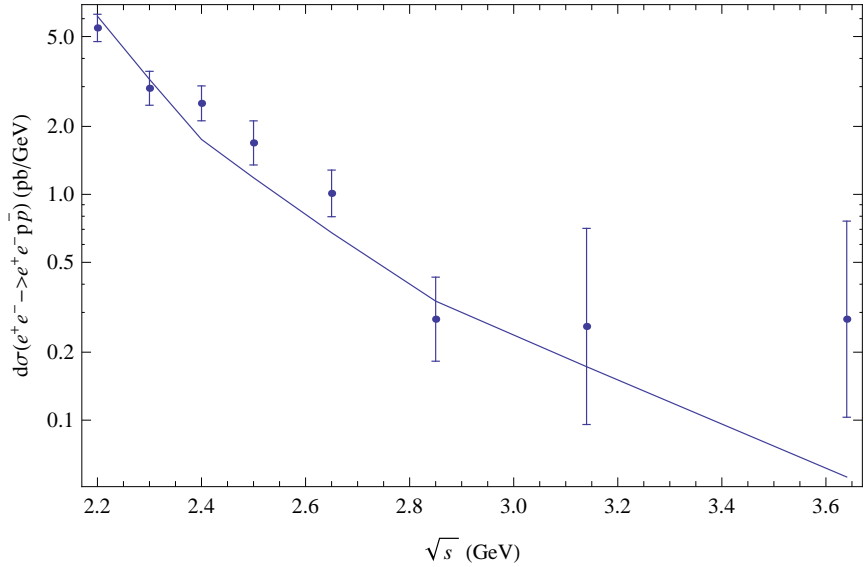}
       \caption{The $e^+ + e^- \to e^+ + e^- + p + \bar{p}$ differential cross section as a function of $\sqrt {s}$
       for the large angle region $|cos\theta^{\ast}| <$ 0.6
       is compared with the experimental data \cite{a13} of the OPAL Collaboration at LEP.}     
       \label{dsigma4}
\end{figure}

In Fig.~\ref{dsigma4}, we illustrated the dependence of the
differential cross section of the process  $e^+ + e^- + p + \bar{p}$
on the center-of-mass energy of the  $p \bar {p}$ system.

It should be noted that as is seen from Fig. 10 the differential
cross section of the process $e^+ + e^- \to e^+ + e^- + p + \bar{p}$
is in satisfactory agreement with the known experimental data of the
OPAL Collaboration at LEP \cite{a13}.

For the calculation of the total cross section of the $e^+ + e^- \to
e^+ + e^- + p + \bar{p}$ process using formulae \eqref{M22} and
\eqref{t}, the integration in \eqref{ds1} over $|cos\theta^{\ast}|
<$ 0.6 and 2.1 GeV $< \sqrt {s_1} <$4.5 GeV is carried out, and
after that, according to \eqref{csee}, we obtain the formula for the
cross section of the $e^+ + e^- \to e^+ + e^- + p + \bar{p}$ process
as a function of $\sqrt {s}$. For energy of $e^+e^-$ collisions in
the L3 Collaboration at LEP the $\sqrt {s}$ dependences are combined
into a single measurement in $<\sqrt {s}>$ = 197 GeV. In this energy
value we have calculated the total cross section of the $e^+ + e^-
\to e^+ + e^- + p + \bar{p}$ process, and we have
\ba
\sigma^{theor}(e^+ + e^- \to e^+ + e^- + p + \bar{p}) = 27.78 \,\,{\rm pb}.
\ea
The experimental data are \cite{a12}
\ba
\sigma^{exp}(e^+ e^- \to e^+ e^- p \bar{p}) = 26.7 \pm 0.9 \pm 2.7 \,\,{\rm pb}.
\ea

\section{Conclusions}\label{conc}

In this paper, we have studied the proton-antiproton production in
the process $e^+e^- \to e^+e^- p \bar {p}$  by the two-photon
mechanism $\gamma\gamma \to p\bar {p}$.

The total cross section of the $e^+e^- \to e^+e^- p \bar {p}$
process was measured in the $p\bar {p}$ in which center-of-mass
energy of the two-photon in the range of 2.1 GeV $< W_{\gamma\gamma}
<$ 4.5 GeV by using the data taken from the L3 detector at $\sqrt
{s_{ee}}$ = 183 -- 209 GeV at LEP.

The differential cross section of the $e^+e^- \to e^+e^- p \bar {p}$
process was measured in the $p\bar {p}$ with center-of-mass energy
of the two-photon in the range of 2.15 GeV $< W_{\gamma\gamma} <$
3.95 GeV using the data taken from the OPAL detector at $\sqrt
{s_{ee}}$ = 183 -- 189 GeV at LEP.

Using the luminosity function, the total cross section of the
process  $\sigma(\gamma\gamma \to p\bar {p})$ as a function of
$W_{\gamma\gamma}$ was obtained from the differential cross section
of the process $d\sigma(e^+e^- \to e^+e^- p \bar {p})/dW$.

We got the master formula for the differential and total cross
section of the $\gamma\gamma \to p\bar {p}$ process. We investigated
the characteristics of the differential and total cross section of
these processes.

The characteristics of the total cross section of the $\gamma\gamma
\to p\bar {p}$ process  in  the energy interval of the two-photon
system $\sqrt {s}$ = 2 GeV $\div$ 6 GeV is studied.  The total cross
section of the $\gamma\gamma \to p\bar {p}$ process was studied for
large angle region $|cos\theta^{\ast}|<$0.6 and the energy of the
two-photon system was changed ($\sqrt {s}$) in the interval 2.1 GeV
$<\sqrt {s} <$ 4.5 GeV.

In this work, the angular distribution of the differential cross
section $\frac{d\sigma (\gamma\gamma \to p\bar
{p})}{dcos\theta^{\ast}}$ of the $\gamma\gamma \to p\bar {p}$
process was studied in detail. Namely, for this, we have separately
considered the angle regions in the interval of
0$<|cos\theta^{\ast}|<$0.6, and regions of  the energies of the
two-photons 2.1 GeV $<\sqrt {s} <$ 2.5 GeV, \,\,2.5 GeV $<\sqrt {s}
<$ 3.0 GeV, \,\,3.0 GeV $<\sqrt {s} <$ 4.5 GeV, respectively.

For full analyzing also we investigated the dependence of the  total
cross section of the $\gamma\gamma \to p\bar {p}$ process on the
energy of two photons in the case of large angle regions
$|cos\theta^{\ast}|<$0.3 and in the angle region
0.3$<|cos\theta^{\ast}|<$0.6.

Moreover, we considered the main process for the proton-antiproton
production in the $e^+e^- \to e^+e^- p \bar {p}$ process and studied
the characteristics of the differential cross section $\frac{d\sigma
(e^+e^- \to e^+e^- p\bar {p})}{dW}$ of this process by the
two-photon mechanism in the large angle regions
$|cos\theta^{\ast}|<$0.6. In Fig.10 we present the dependence  of
the differential cross section   $e^+e^- \to e^+e^- p \bar {p}$
process on the energy of the two-photon  2.15GeV$<W_{\gamma\gamma}<$3.95 GeV .

Also, the total cross section of the $e^+e^- \to e^+e^- p \bar {p}$
process was calculated by using the two-photon mechanism in the
large angle regions $|cos\theta^{\ast}|<$0.6,  at the energy 197 GeV
on the LEP.

All our results  are compared with the experimental data of the L3
and OPAL Collaborations at the LEP \cite{a12,a13}, and they are in
satisfactory agreement.

\section{Acknowledgements}
I am very grateful to Yu. M. Bystritskiy and Egle Tomasi-Gustafsson for the useful discussions.

\end{document}